\documentclass[aps,prl,a4paper,twocolumn,superscriptaddress]{revtex4}

\usepackage{color}
\usepackage{graphicx}
\usepackage{amsmath}
\usepackage{amsfonts}
\usepackage{amssymb}

\definecolor{ngreen}{rgb}{0.2,0.6,0.2}

\definecolor{ngold}{rgb}{0.7,0.6,0.2}

\newcommand{\ket}[1]{\ensuremath{\left| #1 \right\rangle}}

\begin{document}

\title{Optimal multi-photon phase sensing with a single interference fringe}

\author{G.~Y. Xiang}
\affiliation{Centre for Quantum Dynamics and Centre for Quantum Computation and Communication Technology, Griffith University, Brisbane, 4111, Australia}
\affiliation{Key Laboratory of Quantum Information, University of Science and Technology of China, CAS, Hefei 230026, China}
\author{H.~F.~Hofmann}
\affiliation{Graduate School of Advanced Sciences of Matter, Hiroshima University, Kagamiyama 1-3-1, Higashi Hiroshima 739-8530, Japan}
\affiliation{JST,Crest, Sanbancho 5, Chiyoda-ku, Tokyo 102-0075, Japan}
\author{G.~J. Pryde}
\affiliation{Centre for Quantum Dynamics and Centre for Quantum Computation and Communication Technology, Griffith University, Brisbane, 4111, Australia}

\begin{abstract}
Quantum entanglement can help to increase the precision of optical phase measurements beyond the shot noise limit (SNL) to the ultimate Heisenberg limit. However, the $N$-photon parity measurements required to achieve this optimal sensitivity are extremely difficult to realize with current photon detection technologies, requiring high-fidelity resolution of $N+1$ different photon distributions between the output ports. Recent experimental demonstrations of precision beyond the SNL have therefore used only one or two photon-number detection patterns instead of parity measurements. Here we investigate the achievable phase sensitivity of the simple and efficient single interference fringe detection technique. We show that the maximally-entangled ``NOON'' state does \textit{not} achieve optimal phase sensitivity when $N > 4$, rather, we show that the Holland-Burnett state is optimal. We experimentally demonstrate this enhanced sensitivity using a single photon-counted fringe of the six-photon Holland-Burnett state. Specifically, our single-fringe six-photon measurement achieves a phase variance three times below the SNL.
\end{abstract}

\maketitle

Quantum information technologies promise to revolutionize the way we communicate and process information, providing new levels of security~\cite{Gisin2002} and the ability to tackle a range of intractable computational problems~\cite{Ralph2009}. The same physics concepts also provide for schemes to measure~\cite{Giovannetti2004} and manipulate~\cite{Boto2000} the world with precision far beyond that possible with classical techniques.

A particularly important example is the case of optical interferomtery, where information on displacement, velocity, materials properties etc.\ is obtained by detecting a phase shift. Here, photon statistics appear to limit the sensitivities of $N$-photon interferometry to phase uncertainties of $\Delta \phi=1/\sqrt{N}$. However, this shot noise limit (SNL) can be overcome by entangling the photons in a single, fully quantum-coherent state~\cite{Giovannetti2004,Durkin2007}. The ultimate limit of phase sensitivity can then be expressed in terms of an uncertainty relation between the phase $\phi$ and the photon number difference $n_1-n_2$ between the two interferometer paths: $\Delta \phi \Delta(n_1-n_2) \geqslant 1$. For $N$ photons, the maximal uncertainty is $\Delta(n_1-n_2)=N$, resulting in a minimal phase uncertainty $\Delta \phi=1/N$, the Heisenberg limit (HL)~\cite{Giovannetti2004}. The $N$-photon state that achieves this limit is the NOON state $\ket{\psi}_\textrm{NOON}=\tfrac{1}{\sqrt{2}}(\ket{N0}+\ket{0N})$, where the photons are either all in one path or all in the other path~\cite{Dowling2008}. For high $N$, this state describes a quantum superposition of macroscopically distinguishable states. It is therefore not surprising that the generation of NOON states for high $N$ is extremely difficult. To date, optical NOON experiments have only been realized with up to five photons~\cite{Mitchell2004,Walther2004,Nagata2007,Okamoto2008,Gao2010,Afek2010}. Initial methods were based on low efficiency post-selection from down-converted photon pairs, resulting in exponentially poor scaling. Recently, a more favourably-scaling scheme has been demonstrated for postselecting NOON states from entangled states of uncertain photon number~\cite{Afek2010,Ono2008}, but this method is still technically difficult since it requires phase-stabilized interference between two very different light sources. Heralded generation~\cite{Matthews2010} and amplification~\cite{Xiang2010b} of path-entangled states have also been demonstrated, again for small $N$.

Due to the experimental difficulties of generating NOON states, there have also been considerable efforts to exploit the phase sensitivity of states that can be created deterministically from unentangled inputs. As early as 1993, Holland and Burnett pointed out that multiple photon pairs created in parametric down-conversion result in a highly phase sensitive $N$-photon state when the two beams with $N/2$ photons each are injected into each input port of an interferometer~\cite{Holland1993}. Due to photon bunching arising from nonclassical interference, the uncertainty of the photon number difference between the paths inside the interferometer is $\sqrt{N(N+2)/2}$, only about $\sqrt{2}$ lower than the maximal uncertainty of NOON states. For appropriate output measurements, it should therefore be possible to achieve phase uncertainties below $\sqrt{2}/N$, an enhancement over the SNL that scales like the HL as $N$ increases. A 4-photon HB experiment has recently been performed~\cite{Sun08}. Unfortunately, the claim of near-optimal sensitivity in that work was somewhat misleading, since the authors assumed an experimentally urealisitic quantum state fidelity of $100\%$ to define their peak probability, resulting in an overestimate of the phase sensitivity around the peak. Nevertheless, that work suggested that the uncertainty limit of phase resolution might be achieved by measuring only the probability of the equal photon number output \cite{Datta11}. As we show here, this is a significant feature of the HB state that actually makes it more suitable for single fringe phase estimations than the NOON state. \\

\begin{large}
\noindent\textbf{Results}
\end{large}

\noindent \textbf{Single fringe measurements.} Since much of the theory of quantum metrology has focussed on the properties of the input state, it is often implicitly assumed that the optimal measurement strategy for a given input state can be implemented with available technologies. As a result, the technical challenges involved in the experimental realization of the output measurement have not yet received as much attention as the problem of quantum state preparation.  However, an experimental demonstration of phase sensitivity is not complete unless the proper output measurement has been realized. For NOON states of any photon number, the optimal phase sensitivity is obtained from a two-outcome measurement that assigns a value of $+1$ to even photon numbers in the outputs, and $-1$ to odd photon numbers in the output. 
Ideally, this kind of parity measurement could be performed by a quantum circuit that identifies whether the number of photons at one output is odd or even, without actually counting them. At present, it is not known how to realize such a measurement. Practically, therefore, it is necessary to count the precise photon number at one of the outputs to determine if it is odd or even.
In the absence of unit-efficiency photon-number-resolving detectors, this is very difficult to do. 

On the other hand, efficient postselection of a particular photon number can, in principle, be more easily realized with high, but non-unit, efficiency detectors. Exisiting demonstrations with $N>2$ have therefore relied on projecting the output of the interferometer onto just one or two states of definite photon number in each output arm---e.g.\ $\ket{31}$ or $\ket{13}$ in the case of a four-photon experiment~\cite{Okamoto2008}. Here, we will refer to the results of a single projection of this kind as a ``single fringe'', meaning the phase-dependent projection probability for one pattern of photon counting at the interferometer's outputs. Since single fringe measurements test only a single ``yes/no'' condition, the experimental requirements for their realization are much simpler than those for the realization of a detection stage that can separate all $N+1$ photon number distributions in the output. 

Another advantage of single fringe measurements is that the phase sensitivity is directly related to the resolution of the characteristic features in the experimentally determined fringe. Although many experimental demonstrations cite the N-fold increase of the fringe oscillation period as the characteristic feature of path-entangled states, it has recently been noted that the efficiency of the measurement must be taken into account when trying to demonstrate phase sensitivities beyond the SNL~\cite{Resch2007,Okamoto2008}. Consequently, the small number of experiments that have meanwhile demonstrated (generally small) improvements in precision beyond the SNL~\cite{Rarity90,Kuzmich98,Eisenberg05,Nagata2007,Okamoto2008,Sun08,Gao2010,Xiang2010} have focussed on the evaluation of phase sensitivity, and not on the shape of the fringes. Here, our single fringe analysis can help to establish a very general relation between intuitively accessible features of multi-photon interference and fundamental issues of phase sensitivity.

\noindent \textbf{Fisher information analysis}. To understand the precise requirements for phase super-sensitivity in single fringe measurements, it is useful to quantify the amount of phase information that is lost when the remaining $N$ measurement outcomes are lumped together in a single negative outcome. This can be done by expressing phase sensitivity in terms of the Fisher information (FI), denoted $\mathcal{F}$~\cite{Durkin2007}. According to the Cramer-Rao bound~\cite{Cramer1946}, the Fisher information determines the lowest phase uncertainty achievable for a given set of phase dependent measurement probabilities: $\Delta \phi \geqslant 1/\sqrt{\mathcal{F}}$. 
In terms of the $N+1$ photon-counted fringes that correspond to an $N$-photon state, the Fisher information $\mathcal{F}=\mathcal{F}(\phi)$ (and hence the phase sensitivity) is given by
\begin{equation}
\mathcal{F}(\phi)=\sum^{N+1}_{i=1}p_i (\phi) \left(\frac{\partial}{\partial \phi} \ln p_i (\phi)\right)^2
\label{fisherall}
\end{equation}
In quantum metrology, the Fisher information also depends on the measurement strategy~\cite{Braunstein1994}. Fortunately, it can be shown that precise photon counting in the output is an optimal strategy for path-symmetric pure states such as the HB and NOON states~\cite{Hofmann2009}. Therefore, the Fisher information of the  $N+1$ fringes would ideally result in a phase-independent Fisher information $\mathcal{F}=(\Delta(n_1-n_2))^2 $.
For uncorrelated photons, this results in a phase independent sensitivity corresponding to the SNL of $\mathcal{F}=N$, a result that has been confirmed experimentally using photon number resolving detectors~\cite{Pezze2007}. For NOON states, the maximal photon number uncertainty would similarly result in a phase sensitivity at the HL of $\mathcal{F}=N^2$. However, experimental imperfections reduce this ideal value, introducing a phase dependence of the Fisher information, as can be seen in recent results obtained for all 5 fringes of an $N=4$ photon experiment~\cite{Xiang2010}. 

Since a single fringe $i$ contains only a subset of the phase information available in the phase-shifted multi-photon state, the Fisher information $\mathcal{F}_i$ obtained from the single fringe will usually be lower than the complete Fisher information obtained from all $N+1$ fringes. Specifically, the Fisher information of a single fringe is given by the contribution of the fringe $i$ to the sum in Eq.~(\ref{fisherall}), combined with the information from its null fringe:
\begin{eqnarray}
\nonumber \mathcal{F}_i(\phi)&=&p_i (\phi) \left(\frac{\partial}{\partial \phi} \ln p_i (\phi)\right)^2\\
&&+(1-p_i(\phi)) \left(\frac{\partial}{\partial \phi} \ln (1-p_i (\phi))\right)^2
\label{Fishersingle}
\end{eqnarray}
The limited information contained in a subset of photon-counted fringes is one of the reasons that few experiments to date have demonstrated sensitivities below the SNL - most experiments use just one or two of the $N+1$ fringes to obtain an estimate of the phase sensitivity of their states. The success of such experiments depends on the possibility of beating the SNL with just a single measurement fringe, i.e.\ to obtain $\mathcal{F}_i > N$ for a specific measurement setting $i$. It is therefore important to understand the theoretical limits on phase sensitivity imposed by single fringe measurements. 

\noindent \textbf{Optimality of the Holland-Burnett state.} Quantum mechanically, the single fringe probability is determined by $p_i(\phi)=|\langle m_i|\hat{U}(\phi)|\psi\rangle|^2$, where $\hat{U}(\phi)$ is the unitary transformation that describes the phase shift. For a fixed outcome $i$, the roles of measurement outcome $m_i$ and initial state $\psi$ are perfectly symmetric. As a result of this symmetry, the phase sensitivity of single fringe measurements is not only limited by the photon path uncertainty of the initial state $| \psi \rangle$, but also by the corresponding uncertainty in the photon number difference between the paths associated with the output state $| m_i \rangle$. This measurement dependent uncertainty limit can be expressed in terms of the operator $\hat{n}_1-\hat{n}_2$ of photon number difference between the paths inside the interferometer as
\begin{equation}
\label{sflimit}
\mathcal{F}_i \leq \langle m_i |(\hat{n}_1-\hat{n}_2)^2| m_i \rangle,
\end{equation}
where the possible choices of $| m_i \rangle$ depend on the available measurement technologies. 
For photon counting in the output, single fringe measurements cannot exceed a sensitivity given by the path uncertainties of the \textit{detected} photon number states. With $\tfrac{N}{2}+m$ photons in one port and $\tfrac{N}{2}-m$ photons in the other port, this uncertainty limit is equal to $\tfrac{1}{2}N(N+2)-m^2$. The output state with the maximal path uncertainty is the state with equal photon numbers in each port ($m=0$), corresponding to a projective measurement of the HB state. Therefore, the maximal phase sensitivity of single fringe detection is equal to the Fisher information of the HB state, indicating that any higher input state sensitivity will be lost in the measurement. Oppositely, it is relatively easy to show that the $m=0$ fringe of the HB state itself does achieve the maximal phase sensitivity.
(See Methods for details). The HB state is therefore the optimal input state for single fringe measurements, and its Fisher information of $\mathcal{F}_i(0)=F=\tfrac{1}{2}N(N+2)$ is the maximal single fringe Fisher information $\mathcal{F}_i$ for any $N$-photon state. 

\begin{figure}
\center{\includegraphics[width=\columnwidth]{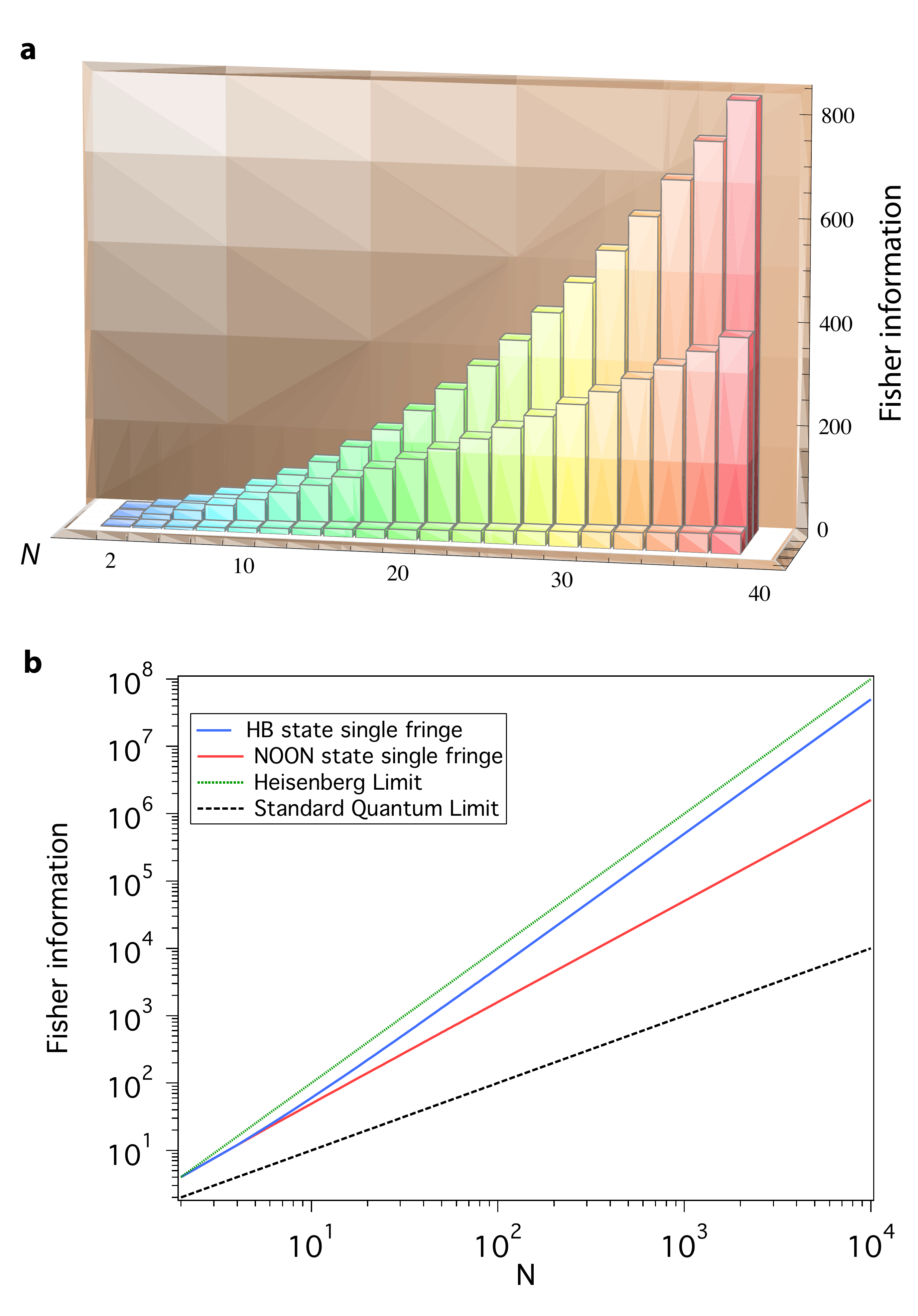}}
\caption{\textbf{Performance of the NOON and HB states with single fringe measurements.} The figures show the Fisher information vs.~$N$ small ($N \leqslant 40$) and large $N$ respectively. Note that HB states are only defined for even $N$. \textbf{a.} Small $N$. Front: SNL; middle: NOON; back:HB. Note that the advantange of the HB state grows as $N$ increases. \textbf{b.} In the asymptotic limit, the single-fringe HB Fisher information scales with the HL, while the NOON state scales worse.}
\label{thyfig}
\end{figure}

By contrast, the NOON state is not optimal for phase estimation with single photon detection fringes, since the selection of a single fringe severely reduces its phase sensitivity. For even photon numbers, the optimal single fringe sensitvity is obtained from the output with equal photon numbers in both ports. However, NOON states only achieve the output uncertainty limit of Eq.~\ref{sflimit} for $N=2$ (where NOON state and HB state are the same state) and for $N=4$. At $N>5$, NOON states actually perform less well than HB states in single fringe measurements,
and do not even achieve the same scaling as the HL, as shown in Fig.~\ref{thyfig}.

\begin{figure}
\center{\includegraphics[width=7.6cm]{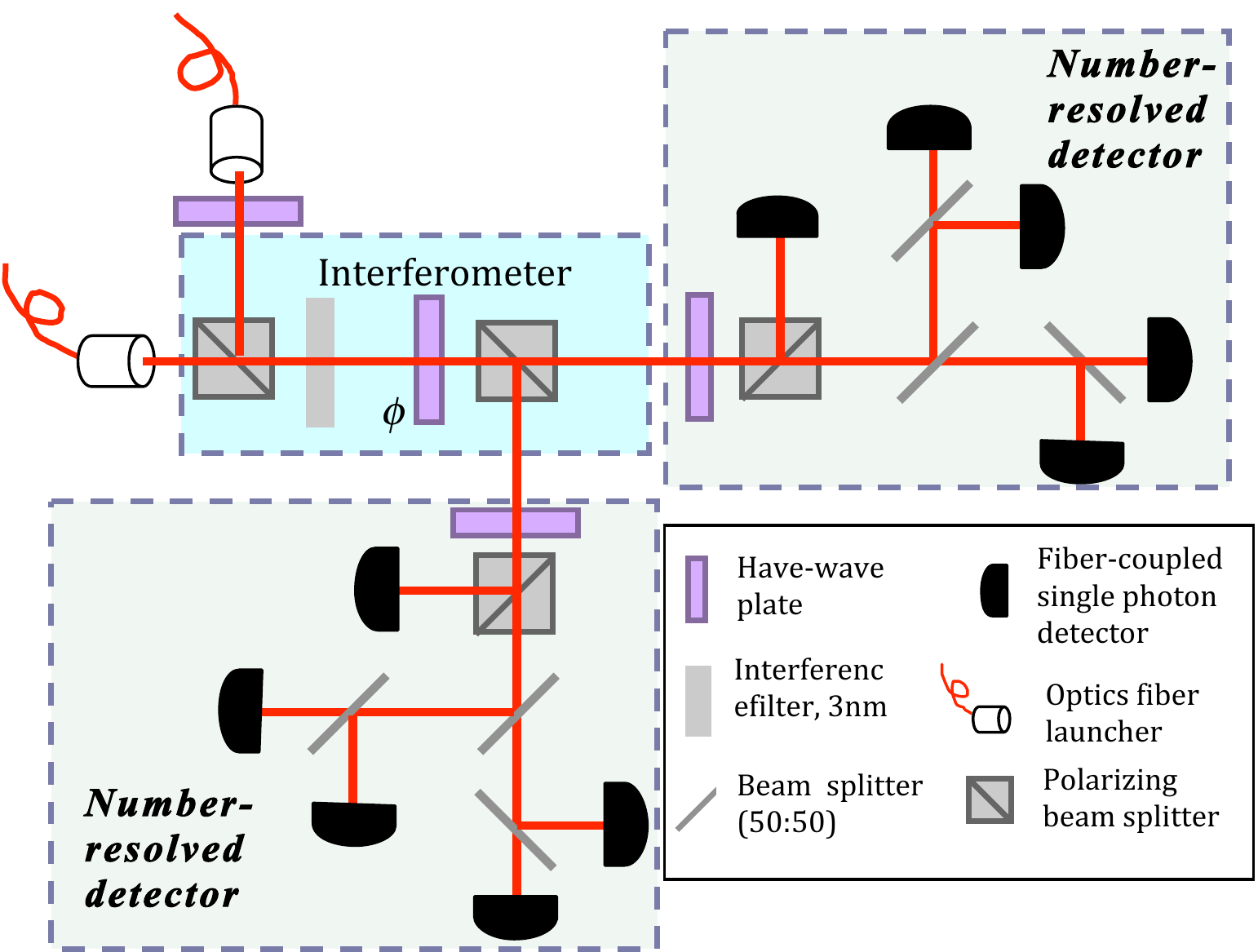}}
\caption{\label{fig:setup}\textbf{Schematic of the experiment.} The input photon pairs are from a spontaneous parametric down-conversion (SPDC) source and biphotons which implement our dual $\tfrac{N}{2}=3$ Fock states, $\ket{\psi}_\textrm{in}=\ket{33}$, are guided to the interferometer with single-mode polarization maintaining optic fibres. These photon states are incident on a polarizing beam splitter and undergo a phase shift between left- and right-circular polarization modes due to the $\phi$ half-wave plate. The final beam splitter recombines the two circularly-polarized interferometric modes, and the output states are measured in the photon number basis by single-photon counting module (SPCM) arrays.}
\label{expsetup}
\end{figure}

\noindent \textbf{Experimental 6-photon single fringe measurements.} We experimentally demonstrate the phase sensitivity obtained using a single-fringe measurement on a six-photon HB state in a polarization-mode interferometer, as shown in Fig.~2. Spontaneous parametric down-conversion supplies the interferometer with pairs of 780~nm triphotons---that is, three indistinguishable photons in each of two spatial modes (see Methods). One horizontally polarized mode and one vertically polarized mode are combined into a single spatial mode using a polarizing beam splitter. A transformation to the right- and left-circular polarization modes is equivalent to the first beam splitter in a Mach-Zehnder interferometer, leading to the Holland-Burnett state
\begin{eqnarray}
\nonumber \ket{3_H 3_V} &=& \frac{1}{4}(\sqrt{5}\ket{6_R 0_L}-\sqrt{3}\ket{4_R 2_L}\\
\nonumber & & \hspace{10pt} +\sqrt{3}\ket{2_R 4_L}-\sqrt{5} \ket{0_R 6_L})
\end{eqnarray}
Thus the right- and left-circular polarization modes of this single spatial mode constitute the arms of the interferometer, and contain the 6-photon entangled states. Phase shifts $\phi$ between these circular polarizations are performed using a half-wave plate with the optic axis at angle $\phi/4$. We implement photon number detection at the outputs of the interferometer by evenly splitting each beam into an array of five single-photon detectors at each output (see Methods). 

We measure the projection onto $\ket{3_H 3_V}$ as $\phi$ is varied, as shown in Fig.~3, representing a measurement of 3 photons at each of the output modes of the interferometer. Theoretically, we expect a probability fringe
\begin{equation}
p_{33}(\phi)=\left(\frac{5}{8} \cos \left[3\phi\right]+\frac{3}{8}\cos \left[\phi\right]\right)^2.
\end{equation}
As shown in Fig.~3, a least-squares fit to a curve of this form has visibility $94 \pm 2 \%$. Here we weight the fit by the poissonian counting error in each data point. From this fitted fringe, we can determine an experimental phase sensitivity, characterized by the Fisher information, Eq.~(\ref{Fishersingle}), and shown in Fig.~4. Although the theoretical optimum FI is achieved at $\phi=0$, the phase information actually goes to zero at that point because of the zero gradient in, and non-unit value of, $p_{33}$ at that phase. Instead, the experimental maximum value of $\mathcal{F}_\textrm{33}^{\textrm{max}}=20.0\pm 0.9$ is found at $\phi^\textrm{max}=15^\circ$. At this point, the phase variance is more than three times smaller than the SNL.

The peak Fisher information significantly exceeds that for the 6-photon SNL, and also exceeds the maximum theoretical single-fringe NOON state FI of $16.91$ for the same visibility. Using several data points in the range $\phi \in [9^\circ,30^\circ]$, we calculated the Fisher information directly from the data, finding $\mathcal{F}_{33}^{\textrm{direct}} (19.6^\circ)=17\pm 5$. This value is consistent with the value obtained from fitting, $\mathcal{F}_{33}^{\textrm{fit}} (19.6^\circ)\approx 19.4$, although the low 6-photon counts leads to a large error in the directly determined value. \\

\begin{figure}
\center{\includegraphics[width=\columnwidth]{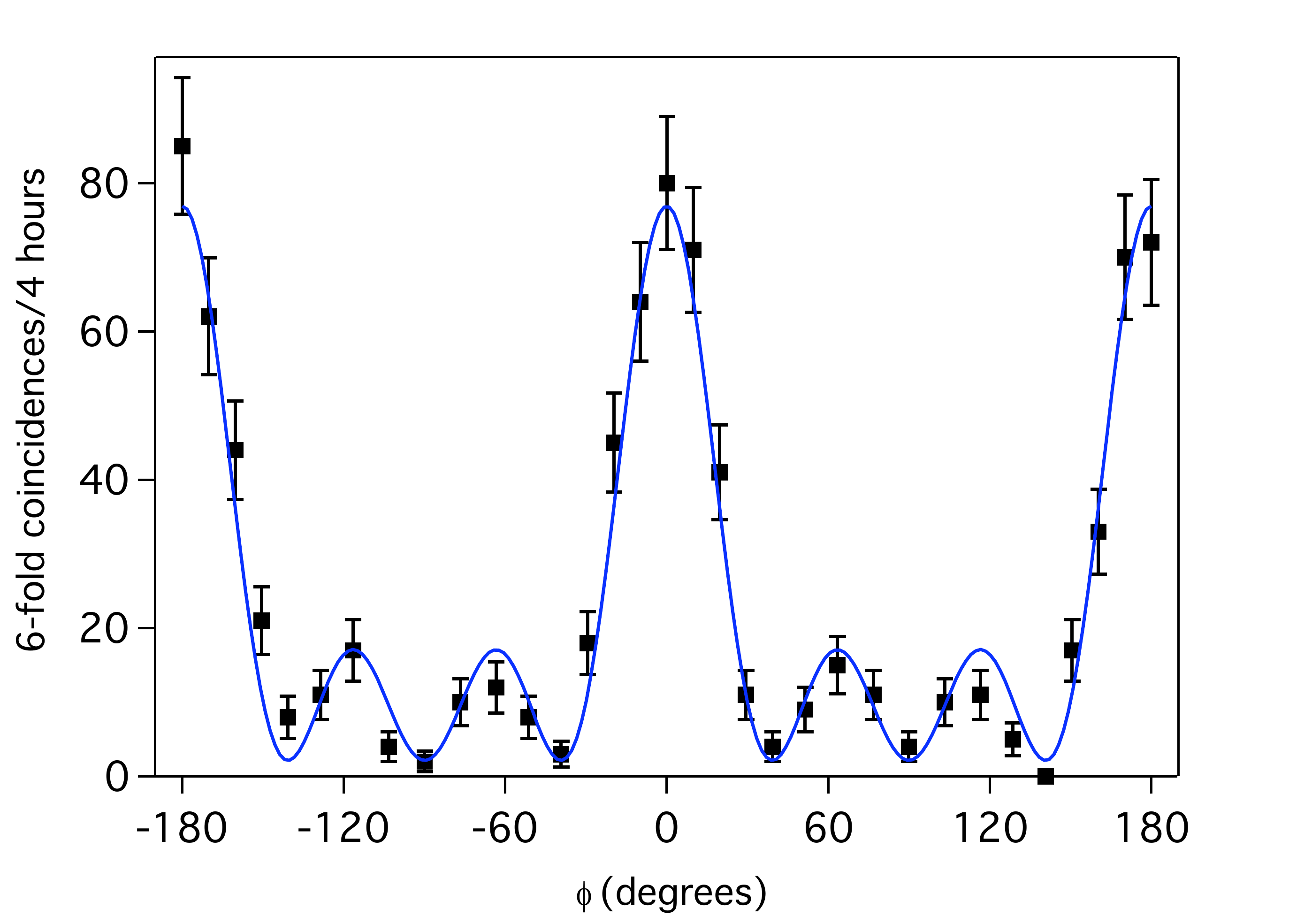}}
\caption{\textbf{Single measurement fringe corresponding to the $\ket{33}$ projection at the output of the interferometer, as the phase shift $\phi$ is varied.} Error bars are derived from poissonian counting statistics. The solid blue curve is a weighted least-squares fit to the fringe, yielding visbility $94\pm 2 \%$.}
\label{datafig}
\end{figure}

\begin{figure}
\center{\includegraphics[width=\columnwidth]{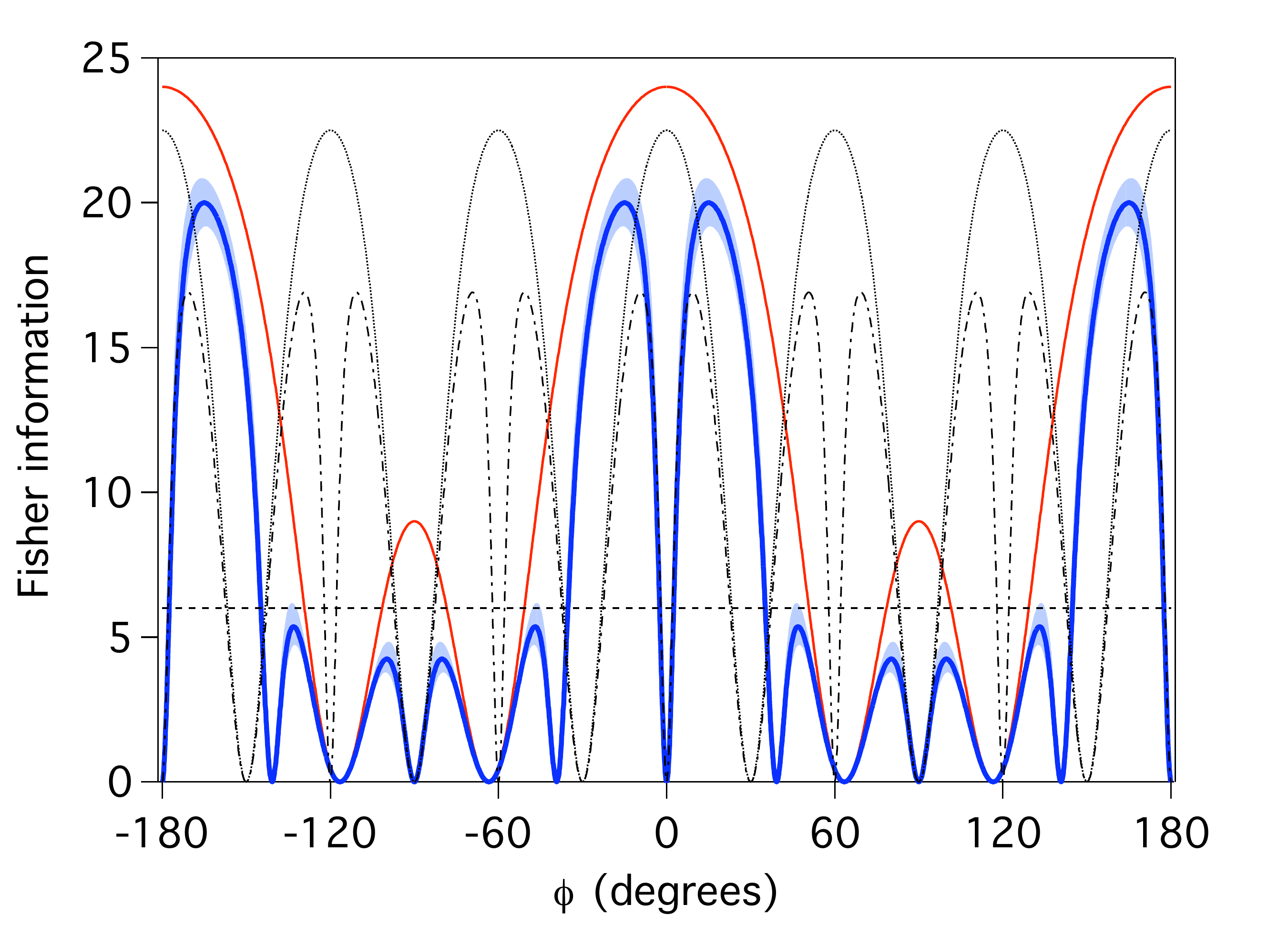}}
\caption{\textbf{Inferred Fisher information.} Fisher information $\mathcal{F}_\textrm{33}$ (solid blue line) calculated from the fit in Fig.~3 for the single-fringe output of our 6-photon Holland-Burnett state. The light blue shading represents one standard deviation of uncertainty, derived from uncertainty in the fit parameters. All uncertainties were derived using standard uncertainty propogation techniques. The solid red curve is the optimal theoretical single-fringe Fisher information for a $\ket{33}$ measurement on the 6-photon HB state. The dashed horizontal line is the SNL, the dotted curve is for a NOON state single fringe with ideal visbility, and the dot-dash curve is for a single-fringe NOON state with visibility equivalent to our experimental data.}
\label{FisherExptFig}
\end{figure}

\begin{large}
\noindent\textbf{Discussion}
\end{large}

\noindent Since single fringe measurements preserve the simplicity of the ``yes/no'' result of single photon detection at arbitrarily high photon numbers, the present system of HB state generation and single fringe detection can be applied to increasing photon numbers without increasing the complexity of the data evaluation. Essentially, the increase in phase sensitivity is directly observed in terms of a sharper central peak in single fringe measurement. HB states therefore provide a more direct and intuitive access to the non-classical enhancement of phase sensitivity by multi-photon entanglement. Since HB states are also easier to generate than NOON states, this simplification of the detection requirements indicates that HB states may be the more reliable option in a wide range of practical applications. While the present realization of HB state generation by spontaneous parametric down-conversion has its limits, the development of bright, high-quality Fock state sources (e.g.\ refs~\cite{Ourjoumtsev2006,Achilles2006,McCusker2009}) may provide the means for preparing high photon number HB states with improved reliability~\cite{ThomasPeter10}. Given suitable Fock state resources and photon detectors with high efficiency (presently under development~\cite{Ralph2009,OBrien2009}), it should be possible to realize phase measurements near the HL with even larger photon numbers, with applications to sensitive optical measurements across the spectrum of science and technology.

We thank B. L. Higgins for assistance with the automated data collection. Part of this work was supported by the Australian Research Council. Part of this work has been supported by the Grant-in-Aid program of the Japanese Society for the Promotion of Science.

\begin{large}
\noindent\textbf{Methods}
\end{large}

\noindent \textbf{Source and detection.} A type-I BBO crystal is pumped by a frequency-doubled mode-locked Ti:Sapphire laser operating at 400~mW average power, with 80~MHz repetition rate and with a pulse length of approximately 150~fs. The spontaneous parametric downconversion outputs from the crystal are coupled to polarization-maintaining optical fibres. These modes are combined into a single spatial mode on a polarizing beam splitter. We restricted the bandwidth of the photons using 3~nm FWHM interference filters.

Each of the two interferometer outputs enters a balanced fan-out array of 5 single photon counting modules, simulating a number-resolving detector. Filtering on detections of 3 photons in each of the interferometer outputs selects out the 6-photon term from the deownconversion source. 

\noindent \textbf{Maximal single fringe phase sensitivity and proof of optimality of the Holland-Burnett state.} A single fringe is given by the probability of a single measurement outcome, $p(\phi)=|\langle m | \psi(\phi) \rangle|^2$. The phase dependence is described by a unitary operation defined by the generator $\hat{h}=(\hat{n}_1-\hat{n}_2)/2$, so that  $\hat{U}=\exp[-i \phi \hat{h}]$. Here, $\hat{n}_1$ and $\hat{n}_2$ are the photon number operators in the paths of the interferometer. The phase derivative of the probability $p(\phi)$ is then given by
\begin{equation}
\frac{\partial}{\partial \phi} p(\phi) = 2 \mathrm{Im}[\langle \psi | m \rangle \langle m | \hat{h} | \psi \rangle].
\end{equation}
The Fisher information of the single fringe is
\begin{equation}
F(\phi) = \frac{\left(\frac{\partial}{\partial \phi} p \right)^2}{p(1-p)}
\leq 4 \frac{|\langle m | \hat{h} | \psi \rangle|^2}{1-p}.
\end{equation}
For all real superpositions of $| m \rangle$ (path-symmetric states), the inequality achieves its bound and becomes an equality. Since $\hat{h}=-i(\hat{a}^\dagger \hat{b}-\hat{b}^\dagger \hat{a})/2$ for output photon number creation and annihilation operators, $\langle m | \hat{h} | m \rangle=0$ and $\hat{h} | m \rangle = -i \Delta h_m | \delta \rangle$, where $\Delta h_m$ is the path uncertanty (or $\hat{h}$ uncertainty) of the detected state $| m \rangle$, and $ | \delta \rangle $ is a state orthogonal to $| m \rangle$. The optimal single fringe resolution is obtained by a superposition of $| m \rangle$ and $| \delta \rangle$, such that
\begin{equation}
F =  4 \frac{\Delta h_m^2 |\langle \delta 
 \psi \rangle|^2}{1-| \langle m | \psi \rangle|^2} = 4 \Delta h_m^2.
\end{equation}
Therefore, the single fringe sensitivity is limited by the generator uncertainty of the output state. In terms of the photon numbers $\hat{n}_1$ and $\hat{n}_2$ in the paths of the interferometer, $h_m^2=4 \langle m | (\hat{n}_1-\hat{n}_2)^2 | m \rangle$, so that the limit of single fringe sensitivity is given by the photon path uncertainty of the output state, as shown in eq. (3) of the paper.  
The maximal path uncertainty of the output is obtained for $m=0$, which corresponds to a projection on the HB state in the output. To actually achieve the output uncertainty limit, the input state must be in a real superposition of this state and the orthogonal state given by $-i \hat{h} | m=0 \rangle$. The straightforward way to meet this requirement is to use the HB state itself as the input. Small phase shifts then produce real superpositions of $| m=0 \rangle$ and $\hat{h} | m=0 \rangle$, ensuring that the maximal phase sensitivity is observed within a sufficiently wide range of phases around $\phi=0$.  

In summary, the discussion above shows that the HB state achieves the maximal phase sensitivity possible with a single photon detection fringe (specifically, the fringe with equal photon numbers at $m=0$). Moreover, this limit can only be achieved by the HB state or by real superpositions of the HB state and the state $-i \hat{h} | m=0 \rangle$, such as the superpositions generated from the initial HB state by small phase shifts. It is therefore impossible to achieve this limit with other states. In particular, NOON states cannot achieve the maximal single fringe sensitivity, because they overlap only partially with the $| m=0 \rangle$ and $\hat{h} | m=0 \rangle$ states at $N>5$. Specifically, the maximal sensitivity of a single NOON state fringe obtained from photon detection is obtained when the output probability $p(m)$ of the fringe is close to zero. This maximal sensitivity is limited by the efficiency $\eta$ that corresponds to the maximal value of the output probability $p(m)$. This probability is equal to two times the symmetric binomial distribution at that point. For equal photon numbers,
\begin{equation}
\frac{F}{N^2} = \frac{N!}{(N/2)!(N/2)!} \left(\frac{1}{2}\right)^{N-1}.
\end{equation} 
In the limit of high photon number, the Sterling formula results in a scaling with $N^{3/2}$ for $F$,
\begin{equation}
F \approx \sqrt{\frac{8}{\pi}} N^{3/2}.
\end{equation}
Therefore the NOON state cannot achieve Heisenberg limited scaling in single fringe inteferometry. Instead, the Fisher information scales with $N^{3/2}$, the geometric mean of SNL and HL.

\end{document}